\documentclass[paper]{revtex4}
\usepackage{epsfig,amsmath,amssymb,color}
\begin{document}
\title{Effect of metal clusters on the
swelling of gold--fluorocarbon--polymer composite films.}

\author{Annalisa Convertino}
\address{Istituto di Chimica dei Materiali C.N.R.--
Area della Ricerca di Roma,\\
 via Salaria Km 29.300, I--00016 Monterotondo St.\ Roma, Italy.}
\author{Antonio Valentini, Anna Bassi}
\address{Istituto Nazionale di Fisica della Materia --
Dipartimento di Fisica -- Universit\`a di Bari,\\
 via Amendola
173, I--70126 Bari, Italy.}
\author{Nicola Cioffi, Luisa Torsi}
\address{Dipartimento di Chimica--Universit\`a di Bari,\\
via Amendola 173, I--70126 Bari, Italy.}
\author{Emilio N.M.\ Cirillo}
\address{Dipartimento Me.\ Mo.\ Mat., Universit\`a di Roma ``La
Sapienza,"\\
via A.\ Scarpa 16, I--00161 Roma, Italy.}
\date{\today}
\begin{abstract}
We have investigated the phenomenon of swelling due to acetone
diffusion in fluorocarbon polymer films doped with different gold
concentrations below the percolation threshold. The presence of
the gold clusters in the polymer is shown to improve the mixing
between the fluorocarbon polymer and the acetone, which is not a
good solvent for this kind of polymers. In order to explain the
experimental results the stoichiometry and the morphology of the
polymer--metal system have been studied and a modified version of
the Flory--Huggins model has been developed.
\end{abstract}
\maketitle 

\narrowtext

Polymers can be combined with inorganic materials to form hybrid
polymer--inorganic structures with a rich variety of physical and
chemical properties. They offer a powerful tool to design novel
devices which take advantage of the immense tailorability and the
low cost processing of polymers. Stacks of alternating layers of
polymer and oxide have been recently used as high reflectivity
dielectric mirrors \cite{[CO0],[FI],[CO1]}, emitting polymers are
sandwiched between reflecting inorganic electrodes in order to
realize organic LED devices \cite{[HI],[TA]}, inorganic--polymer
composite films consisting of inorganic nanoparticles (i.e. metal,
oxide or semiconductor clusters) dispersed in a polymeric matrix
are commonly employed as sensors \cite{[GL],[BR]} and more
recently have been found applications in the fabrication of
all--polymer mirror for the development of all--organic
optoelectronics devices \cite{[CO2],[HO]}.
\par
We study, here, the absorbing behavior of an hybrid
polymer--inorganic structure plunged into a liquid solvent (see
\cite{[EdB],[Da],[DDHP],[D]}). When a polymer is exposed to a
liquid solvent, the liquid enters the polymer and diffuses inside
holes and microvoids in its structure. The polymer swells to an
equilibrium state where the tendency to absorb solvent molecules
is balanced by the elastic response of the network. The {\it
degree of swelling}, namely the volume fraction of polymer in the
mixture at equilibrium, has been calculated in pioneering papers
\cite{[F0]} by means of thermodynamics arguments (see, also,
\cite{[F]}). Many important steps have been recently done toward
the understanding of both the equilibrium (see the recent review
\cite{[EdB]}) and non--equilibrium (see, for instance,
\cite{[RM1],[RM2],[MWJASI],[LP],[TN]}) properties of swelled
polymers. Nevertheless, in spite of the rapidly expanding interest
for pure polymer and hybrid polymer--inorganic structures, due to
the complexity of the swelling phenomena, a complete theory and a
full understanding of the fundamental physics of swelling are
still lacking.
\par
In this letter we study the swelling behavior in a particular
inorganic--polymer system: the gold--fluorocarbon--polymer
composite with a metal content below the percolation threshold.
This  system consists of small gold clusters, with mean diameter
ranging from $2nm$ up to $10nm$ (controlled by the doping),
embedded in the fluorocarbon matrix \cite{[PE],[Na]}. Then,
 we show how the presence of metal clusters can improve
the absorbing power of the fluorocarbon polymer when it is exposed
at the vapors of acetone,
  which is known to be a poor solvent for fluorinated polymers.
Moreover we propose
a genuine entropic interpretation of this effect
based
on a variation on the theme of the Flory--Huggins model \cite{[F0],[F]}.
\par
Samples of gold--fluorcarbon, with typical thickness of about
$100nm$, have been deposited at room temperature on quartz
substrate by ion beam sputtering.
The films have been obtained by cosputtering Teflon and
gold targets with $Ar^{+}$ ion beams.
The $Au$ content $\varphi$, namely the
gold volume fraction in the film, has been varied in the interval
$0\le\varphi\le 0.18$, i.e. below the percolation
threshold which has been found in these materials
to be $\varphi_c\simeq 0.37$ \cite{[PE]}.
More details on the synthesis of these
systems are reported in  \cite{[Qua],[CO2]}.
Transmission Electron Microscopy (TEM) has been employed to assess
the morphology of the composite films. Samples for TEM
characterization have been deposited on 400 mesh Formwar-coated
nickel grids up to a thickness of $10nm$, in order
to be transparent enough to the electron beam.
Film surface composition has been studied by means of X--ray
Photoelectron Spectroscopy (XPS) by using a Leybold LHS10
spectrometer equipped with an unmonochromatised AlK{$\alpha$}
source. Calibration of the spectra has been performed by taking
the $CF_{2}$ component of the C1s electron peak (Binding Energy =
291.6eV) as internal reference \cite{[Shi]}. Curve-fitting has
been performed as reported in reference \cite{[MA]}

In Fig.~1 high and low magnification TEM photographs
of the composite film with $\varphi=0.15$ are reported. As it is
evident from the figure, the composite film shows a high
uniformity and the gold clusters are
equally distributed in all the material.
This allows to exclude the presence in the films
of large area
pores/holes, which did not be expected
in films deposited by means of sputtering technique. Indeed,
the high kinetic energy of the sputtered species produces very dense films.
The clusters of the sample in Fig.~1
are characterized by an average diameter of $3.6 \pm 1.5nm$,
typical value for this kind of materials \cite{[PE],[Na]}.

\par\noindent
A quantitative surface chemical analysis of the films has been
achieved by XPS. Fig.~2 reports a typical high resolution C1s
region for gold--fluorocarbon film ($\varphi=0.15$ in the present
case). The spectrum clearly indicates the presence of different
carbon chemical environments. Six different peaks have been used
to fit the signal. Their attribution is reported in Tab.\ 1, along
with their abundance. The $CF_{3}$, $CF_{2}$, CF, C--CF, C--C in a
fluorinated environment species are typical of a fluorocarbon
matrix \cite{[Qua],[DG]}. The peak falling at $288.5\pm 0.2eV$ can
be simultaneously attributed to carbonyl (C=O) or
fluorinated-unsaturated groups (C=CF) \cite{[DI]}. C=O groups are
minor species that are commonly present in this kind of sample,
that has been exposed to air before ex-situ XPS characterization.
However, the oxygen surface percentage was extremely low in all
the samples (always less than $1\% $),
 and this value is too low as compared to the relatively high intensity
 of the C1s component falling at 288.5eV,
 that should be stoichiometrically balanced by an atomic
 oxygen percentage of at least $3-4\% $.
Consequently, C=CF species are expected to contribute to this
signal. Their presence is confirmed by the weak shoulder (not
fitted) centered at approximately 296eV and due to shake-up
phenomena that are typical of unsaturated carbons \cite{[DI]}. It
is worth noting that the C1s component falling at the lowest
binding energy also has a contribution due to hydrocarbon
contaminants deriving from the vacuum system. Quantitative
treatment of the fitting data has been carried out following two
opposite hypotheses: {\it (i)} the peak is exclusively due to
contaminants, {\it (ii)} the peak is exclusively due to
not-fluorinated branched carbons of the organic film. The results
obtained accordingly have been then averaged. It is important to
note that the metal depositions in/onto fluorocarbon materials are
known \cite{[Shi]} to cause fluorine losses from the polymer, thus
promoting rearrangements and cross--linking processes. The areas
of peaks relevant to tertiary and quaternary carbon species (CF,
C--CF and, eventually, C--C in a fluorinated environment) can be
used to quantify the cross-linking extent (expressed as the
percentage of branched carbons). The results are reported in Tab.
2, where $\Psi_1$ is the percentage of branched C and $\Delta_1$
is the ratio between the branched C percentage and that at
$\varphi=0$ in the hypothesis {\it (i)}, whereas $\Psi_2$ and
$\Delta_2$ are the same quantities calculated in the hypothesis
{\it (ii)}. In the present case the abundance of these components
increases significantly upon gold inclusion and the amount of
branched carbons almost doubled itself in the $\varphi=0.18$
composition.


It has been remarked, see \cite{[BR],[MWJASI]}, that in the
inorganic--polymer composites above the percolation threshold the
absorption power abruptly decreases. Here we consider different
films, with different metal content below the percolation
threshold, and we measure the equilibrium concentration of acetone
in the swelled film as a function of the vapor pressure. We
describe, now, our swelling experiments: the films have been
inserted in a vacuum chamber, characterized by a base pressure
approximately equal to $1\times 10^{-3}mbarr$, where acetone
vapors have been introduced. The pressure $P$ of the acetone
vapors in the vacuum chamber has been measured by a Pirani gauge.
The swelling of the film produces a relative thickness variation
$\Delta t$, that has been monitored by an ellipsometer operating
at the single wavelength of $632.8nm$. In the range of the acetone
vapor pressure analyzed we did not observe any change in the
refractive index for the sample with $\varphi=0$.
  Then this allows us to express  the volume
fraction $y$ of solvent in the mixture as the equality $y=\Delta
t/(t+\Delta t)$ and to obtain $y$
 as a function of the relative
pressure $P/P_0$ (where $P_0=242mbarr$ is the saturation vapor
pressure of the pure acetone at room temperature \cite{[Ri]}) by
only measuring the thickness variations. In Fig.~3 our results
have been plotted for different $Au$ contents, namely
$\varphi=0\;(\bullet),\;0.05\;(\blacksquare),\;0.13\;
(\blacktriangle),\;0.18\;(\blacktriangledown)$ (the solid lines
are only guides for eyes). The sample with $\varphi=0$ shows a
very low absorbing power, which saturates at $P/P_0\le 0.05$. This
behavior confirms the low solubility of the fluorocarbon polymer
in acetone. From Fig.~3 it is evident that the swelling effect is
enhanced by the presence of the gold clusters in the polymer,
indeed at fixed values of $P/P_{0}$ the volume fraction of solvent
$y$ increases with the $Au$ content.

This effect is rather surprising because the presence of the metal
should reduce the volume effectively occupied by the polymer in
the sample and inhibit the absorption. Moreover the XPS
measurements indicate that the inclusion of metal clusters
increases the cross-linking in the films, which become more rigid
and less soluble \cite{[F0]}, and the TEM results exclude the
presence of holes and pores in our films. This phenomenon looks
definitely interesting both from the point of view of the
applications and from that of the understanding of the fundamental
physics of swelling. Indeed, it suggests that the presence of
topological defects in the polymer matrix, such as our
non--percolating golden obstacles, favors the absorption
phenomenon. We will try to explain our results via an entropy
based argument.

The equilibrium concentration $y$ of solvent in the mixture is the
solution of the equation:
\begin{equation}
RT\log\frac{P}{P_0}=
\frac{\partial}{\partial n}\Delta G(y)
\;\;\; ,
\label{eq:equilibrio}
\end{equation}
where $R$ is the gas constant, $T$ is the temperature, $P_0$ is
the saturation vapor pressure of the pure solvent, $n$ is the
number of moles of solvent in the mixture and $\Delta G(y)$ is the
free energy of mixing. The left--hand side is nothing but the
difference between the chemical potential of the pure solvent in
the vapor and liquid phase, whereas the right--hand term is the
difference between the chemical potential of the solvent in the
mixture and in the pure liquid phase. Flory--Huggins (see
\cite{[F0],[F]}) derives the expression of the derivative of the
free energy of mixing $\partial\Delta
G^{{\mathrm{pure}}}(y)/\partial n= RT\left\{ \alpha w^2+\log
y+w+\beta\left[w^{1/3}-w/2\right]\right\}$ where $w=1-y$ is the
volume fraction of polymer in the mixture, the {\it degree of
swelling}, and $\alpha,\beta$ are dimensionless positive constants
respectively related to the heat of dilution and to the effect of
the cross--links. For a suitable choice of the parameters $\alpha$
and $\beta$, the above function increases monotonically from
$-\infty$ to $0$ when $0\le y\le 1$. Hence, the polymer absorbs
solvent until $y$ is so large that equation (\ref{eq:equilibrio})
is satisfied.
\par
The fact that the chemical potential of the solvent in the
mixture increases with $y$ can be understood
through an heuristic argument: the free energy of mixing can
be written as $\Delta G=\Delta H -T\Delta S$, where
$\Delta H$ and $\Delta S$ are respectively the enthalpy and
the entropy of mixing. Now,
$\partial\Delta S/\partial n$ is surely positive,
but decreasing with $y$. Indeed, when more solvent is added to
the mixture its entropy increases, but this increment
is smaller when $y$ is larger.
\par
In the case of a inorganic--polymer composite it is natural to
suppose that the entropy of mixing is more sensitive to the
absorption of solvent. So we can conjecture that $\partial\Delta
S^{{\mathrm{pure}}}/\partial n \ll
\partial\Delta S^{{\mathrm{comp}}}/\partial n$: this implies that
the chemical potential is smaller in the case of composite materials,
so the equilibrium (\ref{eq:equilibrio}) is reached for larger values of
$y$.
A modified version of the lattice Flory--Huggins model leads to
$
\partial\Delta G^{{\mathrm{comp}}}(y)\partial n=
\partial\Delta G^{{\mathrm{pure}}}(y)\partial n
-\lambda\varphi w^2/(1-\lambda\varphi w)$,
where $w=(1-y)(1-\varphi)$ is the degree of swelling, $\lambda$ is
a dimensionless positive constant and the new term
$-\lambda\varphi w^2/(1-\lambda\varphi w)$ is the entropic
correction conjectured above. In Fig.~4 our data have been fitted
with the parameters $\alpha=0.8$,
$\beta=0.1\times(\Delta_1+\Delta_2)/2$ and $\lambda=6$. The low
pressure ($P/P_0\le 0.05$) experimental data are quite well
reproduced by equation (\ref{eq:equilibrio}) with the corrected
chemical potential and $\varphi$ assuming the measured values
$\varphi=0,0.05,0.13,0.18$. It is important to remark that in the
four graphs in Fig.~4 the parameters $\alpha$ and $\lambda$ are
kept fixed, and only the parameter $\beta$ and $\varphi$ are
varied; this suggests that the effect is mainly entropic.

At $P/P_0=0.05$ the trend of the experimental data changes
abruptly, as a consequence of the insolubility of the fluorocarbon
polymer matrix in acetone, indeed at $\varphi=0$ a saturation
solvent concentration is approached. Flory--Huggins models cannot
describe the saturation regime, but as $\varphi$ increases the
composite films become more and more soluble and the agreement
between experimental data a theoretical predictions is improved.
Although this agreement is reasonably good at $\varphi\not= 0$, it
is clear that different swelling mechanism should be taken into
account. A possible explanation could be the electric and chemical
interaction between acetone molecules and the gold clusters.

Finally, we briefly describe our modified Flory--Huggins lattice
model: a three--dimensional cubic lattice with $N$ sites is filled
with $N_w$ polymer chains, each chain is made of $\kappa$ monomers
and each monomer occupies a lattice site; the remaining $N-\kappa
N_w$ sites are occupied by $N_y=N-\kappa N_w$ molecules of
solvent. The support of each chain is a sequence of $\kappa$
pairwise nearest neighboring sites. The effect of the doping is
introduced by supposing that some of the bonds on the lattice are
broken, in the sense that two neighboring monomers of a chain
cannot occupy two neighboring sites of the lattice sharing a
broken bond. The density of broken bonds is chosen equal to
$\lambda\kappa N_w\varphi/N$ where $\lambda$ is a positive
constant. The configurational entropy is then computed as the
Boltzmann constant times the logarithm of the number of possible
configurations of the system. By deriving the entropy of mixing
with respect to $N_y$ we obtain our entropic correction term.
\par
We conclude  that  in gold--fluorocarbon--polymer composite films
with metal content below the percolation threshold the swelling
phenomenon is enhanced by the presence of inorganic clusters in
the polymer. These results suggest that in the swelling of hybrid
inorganic--organic materials, like inorganic--polymer composites,
the inorganic component plays an important role in the swelling of
the polymeric part so that a rather extensive territory could
still be explored if different inorganic particles and matrix
polymers were considered. This could improve the understanding of
an important phenomenon involving organic matter and would also
offer a powerful tool in the design and in the fabrication of
polymer devices employing the swelling.

\acknowledgments

\noindent Dr. T. Beve--Zacheo, who performed the TEM analysis, is
gratefully acknowledged.
 E.C.\ wishes to express his thanks to
the European network ``Stochastic Analysis and its Applications"
ERB--FMRX--CT96--0075 for financial support.

\vskip 0.5 cm
\par\noindent
\begin{center}
{\large\bf Figure Captions}
\end{center}

\par\noindent
{\large\bf Fig.~1:} TEM photographs relevant to a
gold--fluorocarbon film having a gold volumetric fraction
$\varphi=0.15$ and a thickness of $10nm$. The gold cores mean
diameter was $3.6\pm 1.5 nm$.

\vskip 0.3 cm
\par\noindent
{\large\bf Fig.~2:} High resolution C1s XPS spectrum and
curve--fit results relevant to gold--fluorocarbon film
$\varphi=0.15$.

\vskip 0.3 cm
\par\noindent
{\large\bf Fig.~3:} The volume fraction of solvent in the mixture
as a function of the relative pressure $P/P_0$ at different $Au$
contents: $\varphi=0\;(\bullet),\;0.05\;(\blacksquare),\;0.13\;
(\blacktriangle),\;0.18\;(\blacktriangledown)$ (the solid lines
are only guides for eyes).

\vskip 0.3 cm
\par\noindent
{\large\bf Fig.~4:}
Comparison between experimental data and theoretical results (solid lines)
for the volume fraction of solvent in the mixture at different $Au$ contents:
a) $\varphi=0$;, b) $\varphi=0.05$; c) $\varphi=0.13$;
d) $\varphi=0.18$.

\newpage
\widetext
\begin{center}
{\large\bf Tables}
\end{center}

\vskip 1.5 cm
\begin{center}
\begin{tabular}{c|c|c|c|c|c|c}
\hline\hline
 & $CF_3$ & $CF_2$ & $CF$ & $C=O$ and $C=CF$ & $C$--$CF$ &$C$--$C$\\
 \hline\hline Binding energy (eV) & $293.7\pm
0.1$ & $291.8\pm 0.1$ &
                      $289.8\pm 0.1$ & $288.5\pm 0.2$ &
                      $286.9\pm 0.3$ & $285.2\pm 0.3$ \\
\hline
\%                  & $10.6\pm 0.3$ & $43.2\pm 0.3$ &
                      $16.5\pm 0.7$ & $9.4\pm 0.7$  &
                      $12.8\pm 0.5$ & $7.4\pm 0.4$  \\
\hline\hline
\end{tabular}
\vskip 0.5 cm
{\bf Tab.\ 1:} Peak attributions and quantitative
data to the fitting reported in Fig.~2.
\end{center}

\vskip 1.5 cm
\begin{center}
\begin{tabular}{c|c|c|c|c}
\hline\hline
 & $\varphi=0$ & $\varphi=0.05$ & $\varphi=0.13$ & $\varphi=0.18$ \\
\hline\hline
$\Psi_1$ & 16.0 & 25.0 & 30.1 & 31.6 \\
\hline
$\Delta_1$ &1 & 1.6& 1.9 & 2 \\
\hline
$\Psi_2$ & 16.6 & 27.4 &
34.8 & 37.3 \\
\hline
$\Delta_2$ & 1 & 1.6 & 2.1 & 2.2 \\
\hline
$(\Delta_1+\Delta_2)/2$ & 1 & 1.6 & 2.0 & 2.1 \\
\hline\hline
\end{tabular}
\vskip 0.5 cm

{\bf Tab.\ 2:} Estimations of the cross--linking extent upon
varying the gold content in the composite films. $\Psi_1$ is the
percentage of branched C and $\Delta_1$ is the ratio between the
branched C percentage and that at $\varphi=0$ in the hypothesis
{\it (i)}, whereas $\Psi_2$ and $\Delta_2$ are the same quantities
calculated in the hypothesis {\it (ii)}.

\end{center}

\newpage
\vskip +20mm
\includegraphics[]{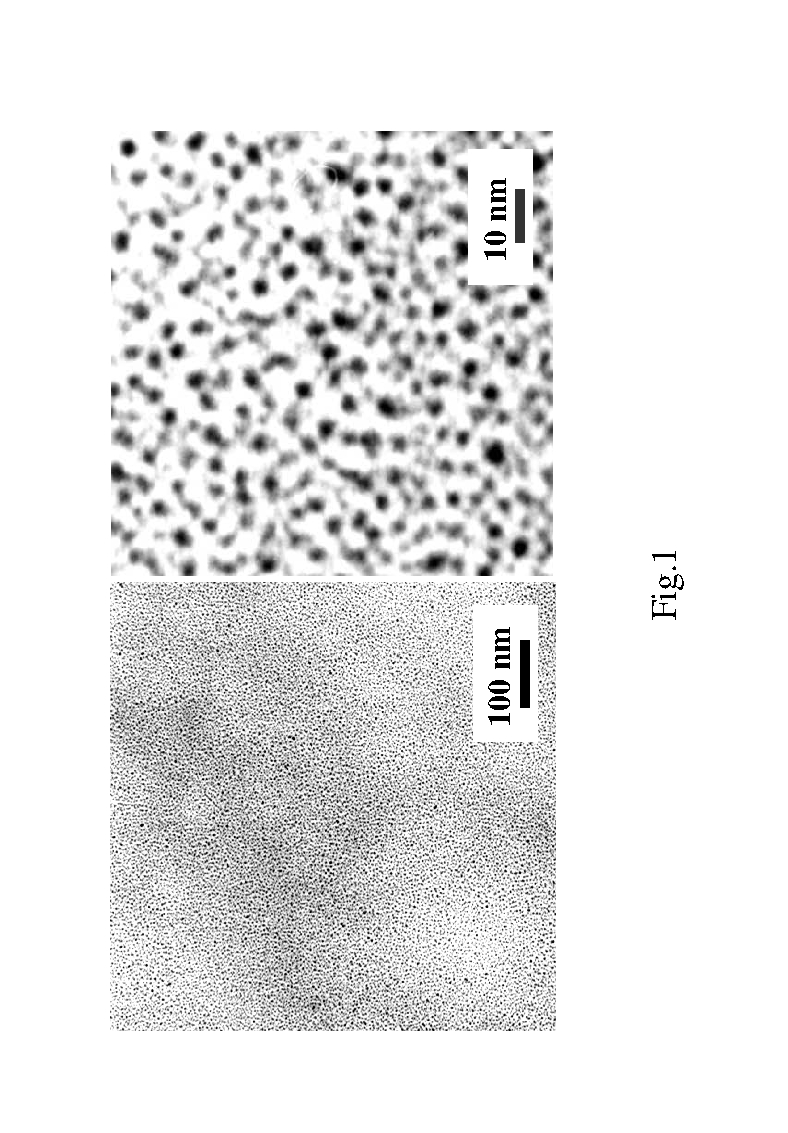}
\vskip +10mm

\vskip +20mm
\includegraphics[]{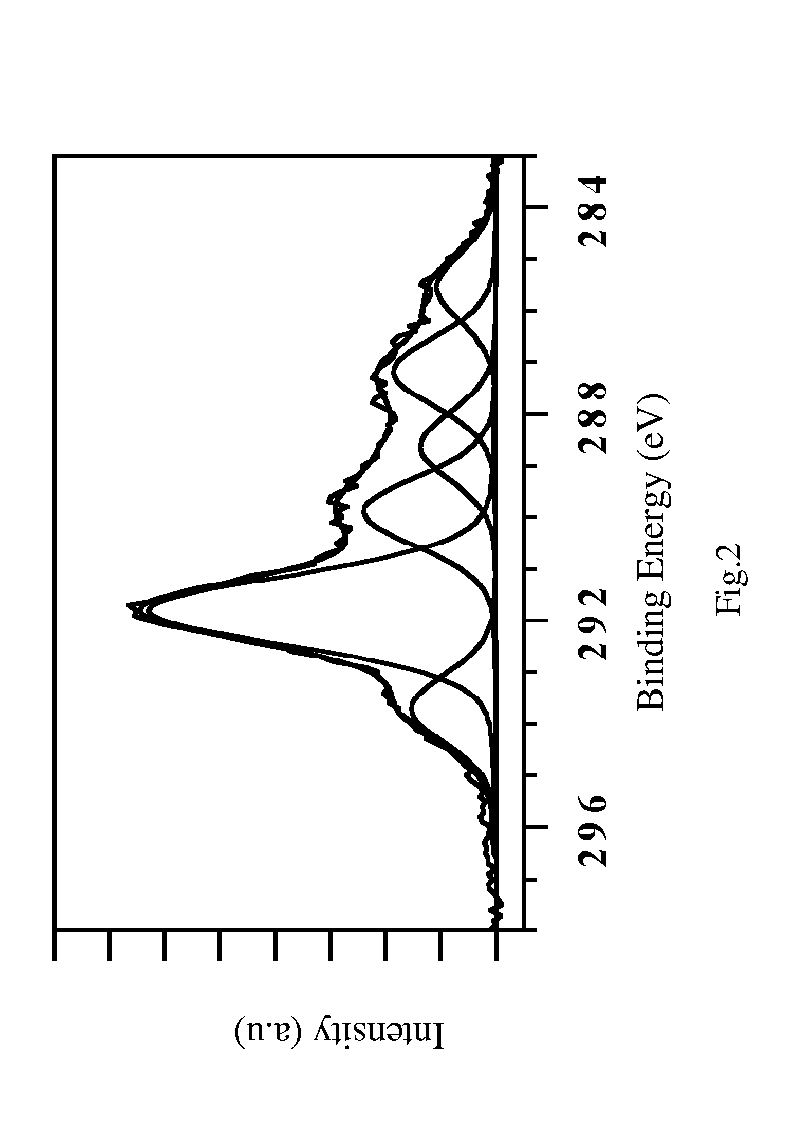}
\vskip +10mm

\vskip +20mm
\includegraphics[]{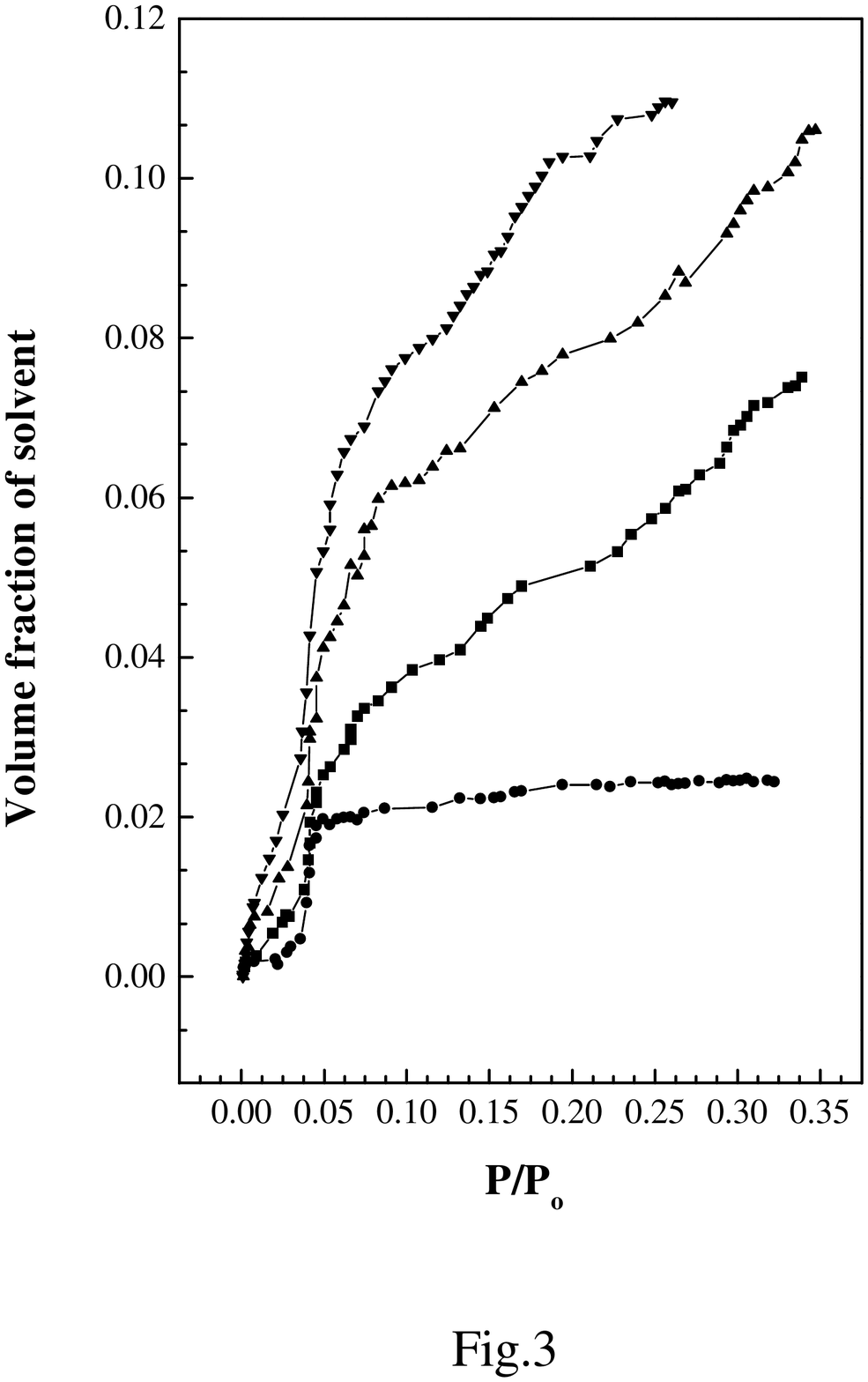}
\vskip +10mm

\vskip +20mm
\includegraphics[angle=90]{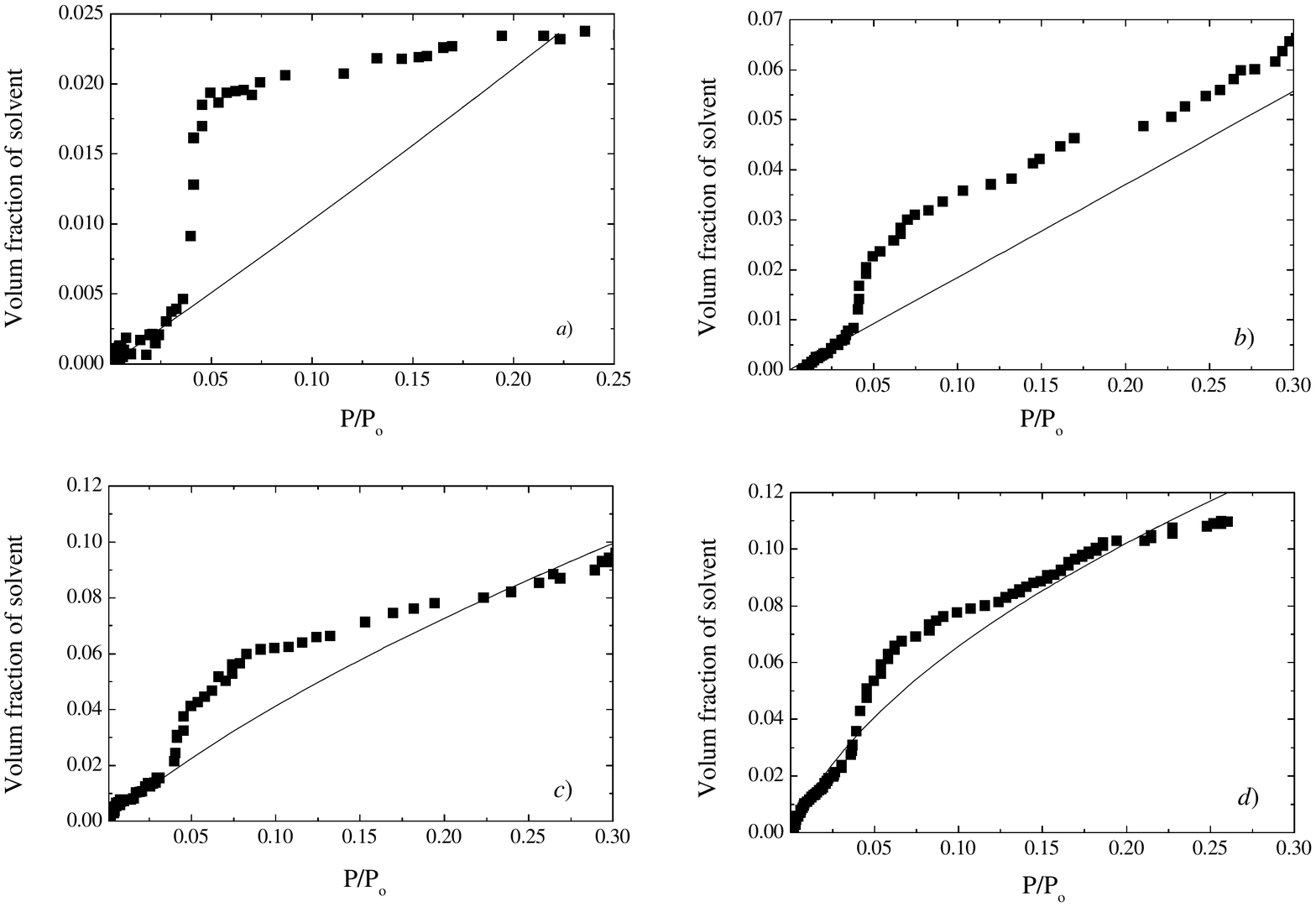}
\vskip +10mm

\end{document}